%% file: 3dph.tex
\documentclass[prl,aps,twocolumn]{revtex4-1}%

\usepackage{amsmath}
\usepackage[utf8]{inputenc}
\usepackage{graphicx} 
\usepackage{amssymb}
\usepackage{verbatim}
\usepackage[colorlinks=true, pdfstartview=FitV, linkcolor=blue, citecolor=blue, urlcolor=blue]{hyperref} 
\usepackage{pifont}
\usepackage{rotating}

\def\C        {{$^{13}$C \/}}

\def\eg       {{\it e.g.}}

\newcommand{\Cn}[1]{$^{13}$C$_{#1}$}

\newcommand{\mr}[1]{\mathrm{#1}}
\newcommand{\unit}[1]{\,\mathrm{#1}}

\newcommand{\us}{\,\mu{\rm s}}
\newcommand{\uT}{\,\mu{\rm T}}

\newcommand{\yn}{\gamma_\mr{n}}
\newcommand{\angstrom}{\textup{\AA}}

\newcommand{\Ia}{\hat{I}_{\vec{a}}}

\newcommand{\apar}{a_{||}}
\newcommand{\aperp}{a_{\perp}}

\newcommand{\Bo}{B_0}

\newcommand{\mS}{m_S}

\newcommand{\Dphi}{\Delta\phi}
\newcommand{\phicoil}{\phi_\mr{coil}}
\newcommand{\phirf}{\phi_\mr{rf}}
\newcommand{\pihalf}{\pi/2}

\newcommand{\veca}{\vec{a}}
\newcommand{\vecAz}{\vec{A}_z}
\newcommand{\vecBcoil}{\vec{B}_\mr{coil}}

\newcommand{\vecr}{\vec{r}}

\begin{document}

\title{Three-dimensional nuclear spin positioning using coherent radio-frequency control}

\author{J. Zopes$^\dagger$, K. Herb$^\dagger$, K. S. Cujia$^\dagger$, C. L. Degen$^\ast$}
\affiliation{Department of Physics, ETH Zurich, Otto Stern Weg 1, 8093 Zurich, Switzerland.}

\begin{abstract}
Distance measurements via the dipolar interaction are fundamental to the application of nuclear magnetic resonance (NMR) to molecular structure determination, but they only provide information on the absolute distance $r$ and polar angle $\theta$ between spins.  In this Letter, we present a protocol to also retrieve the azimuth angle $\phi$.  Our method relies on measuring the nuclear precession phase after application of a control pulse with a calibrated external radio-frequency coil.  We experimentally demonstrate three-dimensional positioning of individual \C nuclear spins in a diamond host crystal relative to the central electronic spin of a single nitrogen-vacancy center.  The ability to pinpoint three-dimensional nuclear locations is central for realizing a nanoscale NMR technique that can image the structure of single molecules with atomic resolution.
\end{abstract}


\maketitle


Nuclear magnetic resonance (NMR) and electron paramagnetic resonance (EPR) spectroscopy are among the most important analytical methods in structural biology and the chemical sciences.
By combining local chemical information of atoms with pair-wise distance constraints, it becomes possible to reconstruct three-dimensional structures or structural changes of proteins and other biomolecules.
While conventional NMR typically analyzes large ensembles of molecules, considerable effort has recently been expended on advancing NMR detection to the level of individual molecules \cite{poggio10,wrachtrup16,schwartz17}.  If successfully extended to the atomic scale, NMR could enable direct imaging of three-dimensional molecular structures, with many applications in structural biology and the nanosciences.  A promising platform for this task are diamond chips containing near-surface nitrogen-vacancy (NV) centers whose electronic spins can be exploited as sensitive local NMR probes \cite{staudacher13,mamin13}.

Structural imaging of single molecules involves determining the three-dimensional coordinates and elemental species of the constituent nuclei.  In NV-NMR, information on the spatial position can be gained from the dipolar part of the hyperfine interaction between the nuclei and the central electronic spin \cite{taminiau12,kolkowitz12,zhao12}.  Because of the axial symmetry of the dipolar interaction, however, only the absolute distance $r$ and the polar (inter-spin) angle $\theta$ can be inferred from a NMR spectroscopy measurement.  Although the axial symmetry can be broken by a static \cite{zhao12} or dynamic \cite{zopes18} transverse magnetic field, determination of the azimuth angle $\phi$, required for reconstructing the full three-dimensional distance vector $\vecr = (r,\theta,\phi)$, has remained challenging \cite{laraoui15prb,wang16,sasaki18}.

\begin{figure}[t!]
\includegraphics[width=0.45\textwidth]{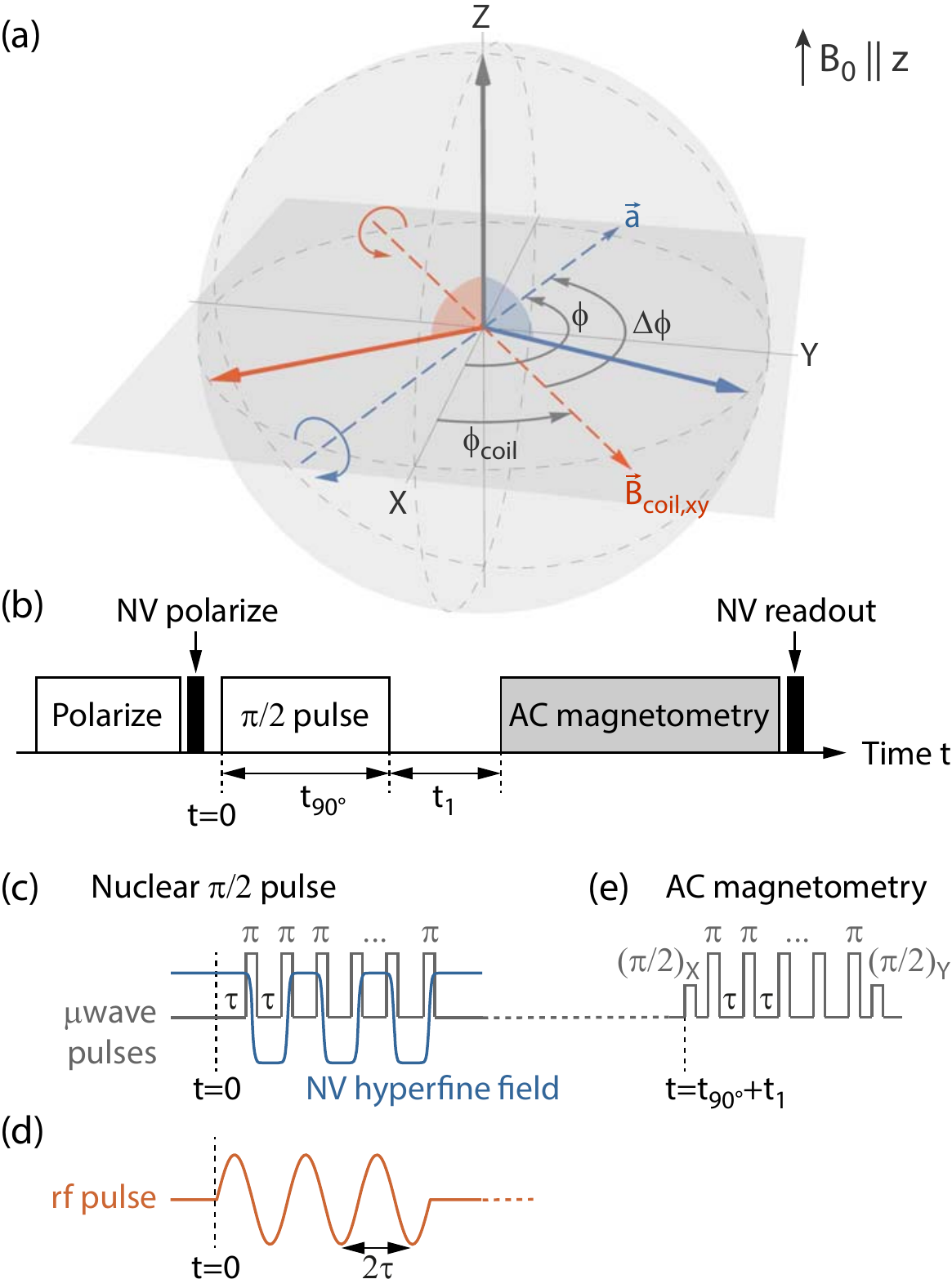}
\caption{(a) Bloch-sphere schematic of a nuclear spin before (grey arrow) and after (colored arrows) application of a $\pihalf$ rotation.
The rotation is either mediated by the hyperfine interaction (blue-dashed axis) or a radio-frequency pulse generated by an external micro-coil (orange-dashed axis).
The different azimuth angles of the rotation axes are translated into a phase difference $\Dphi$ of the nuclear spin precession,
thereby linking the known orientation of the coil field to the \textit{a priori} unknown azimuth orientation of the inter-spin vector.  
(b) Pulse sequence used to measure the phase of the nuclear spin precession.
The nuclear $\pihalf$ pulse is implemented either (c) by a modulation of the NV center's hyperfine field using periodic microwave $\pi$ pulses or (d) by driving with an external rf coil.  The modulation frequency $1/(2\tau)$ is matched to the resonance of the nuclear spin.
(e) AC magnetometry is implemented by a Carr-Purcell-Meiboom-Gill sequence of microwave pulses.  The sequence maps the nuclear component $\langle\Ia\rangle$ that is parallel to the hyperfine axis $\veca\propto(\cos\phi,\sin\phi)$ onto the optically detectable polarization state of the NV center.  To register the nuclear precession we sample $\langle\Ia\rangle$ for a series of waiting times $t_1$.
}
\label{fig:fig1}
\end{figure}
%


In this Letter, we demonstrate a simple and precise method for retrieving the azimuth $\phi$ of the inter-spin vector, allowing us to perform full three-dimensional nuclear distance measurements.  Our technique relies on measuring the nuclear precession phase after application of a radio-frequency (rf) pulse by an external micro-coil.  We determine $\phi$ at low and high magnetic fields, and for polarized as well as unpolarized nuclear spins.  We exemplify our method by mapping the three-dimensional locations of \C nuclei for distances up to $11 \unit{\angstrom}$ and angular uncertainties below $4^\circ$.



Our scheme for measuring the azimuth angle is introduced in Fig. \ref{fig:fig1}(a-d): starting from a polarized nuclear state, we perform a $\pihalf$ rotation of the nuclear spin.  The rotation is generated either by modulating the hyperfine field of the NV center using microwave pulses (Fig. \ref{fig:fig1}c), or by applying a rf pulse with an external coil (Fig. \ref{fig:fig1}d).  Subsequently, we let the nuclear spin precess in the equatorial plane of the Bloch sphere and detect the frequency and phase of the precession by an AC magnetometry measurement with the NV center \cite{taylor08,delange11,kotler11} (Fig. \ref{fig:fig1}e).

Crucially, the starting phase of the nuclear precession at $t_1=0$ is set by the axis of the $\pihalf$ rotation, which is determined by the spatial direction of the rf field in the laboratory frame of reference.  When driving the nuclear rotation via the hyperfine interaction, the rf field direction is given by $\vecAz/\yn$, where $\vecAz = (\aperp\cos\phi,\aperp\sin\phi,\apar)$ is the secular part of the hyperfine tensor, $\apar$ and $\aperp$ are the parallel and transverse hyperfine coupling parameters \cite{taminiau14,boss16}, and $\yn$ is the nuclear gyromagnetic ratio (Fig. \ref{fig:fig1}a, blue).  Conversely, if the external coil is used to generate the rf field, the rotation axis is given by the in-plane component of the coil field $\vecBcoil$ (Fig. \ref{fig:fig1}a, red).  By comparing the phases of the precession signals, we directly obtain the relative angle $\Dphi$ between the unknown orientation of the hyperfine vector $\phi$ and the calibrated orientation $\phicoil$ of the external coil field.


We experimentally determine the $\phi$ angles of three \C nuclear spins from three different NV centers in two single-crystal diamond chips.  We optically polarize and read out the NV spin by short laser pulses ($\sim 2\unit{\us}$) and detect the fluorescence intensity in a confocal microscope arrangement.  Microwave pulses at $\sim 2.5\unit{GHz}$ are used to actuate the $\mS=0 \leftrightarrow \mS=-1$ electronic spin transition.  To polarize the nuclear spins, we transfer polarization from the optically aligned NV center using dynamic nuclear polarization with a repetitive initialization sequence \cite{taminiau14,cujia18}. AC magnetometry is performed by a periodic sequence of microwave $\pi$ pulses with XY8 phase cycling \cite{gullion90} enclosed by two $\pi/2$ pulses that are phase-shifted by $90^\circ$ \cite{taylor08,boss17}.  We use a permanent magnet to apply bias fields of $\Bo\sim 10\unit{mT}$ and $200\unit{mT}$ for low field and high field experiments, respectively, aligned to within $1^{\circ}$ of the NV quantization axis.

\begin{figure}[]
	\includegraphics[width=0.48\textwidth]{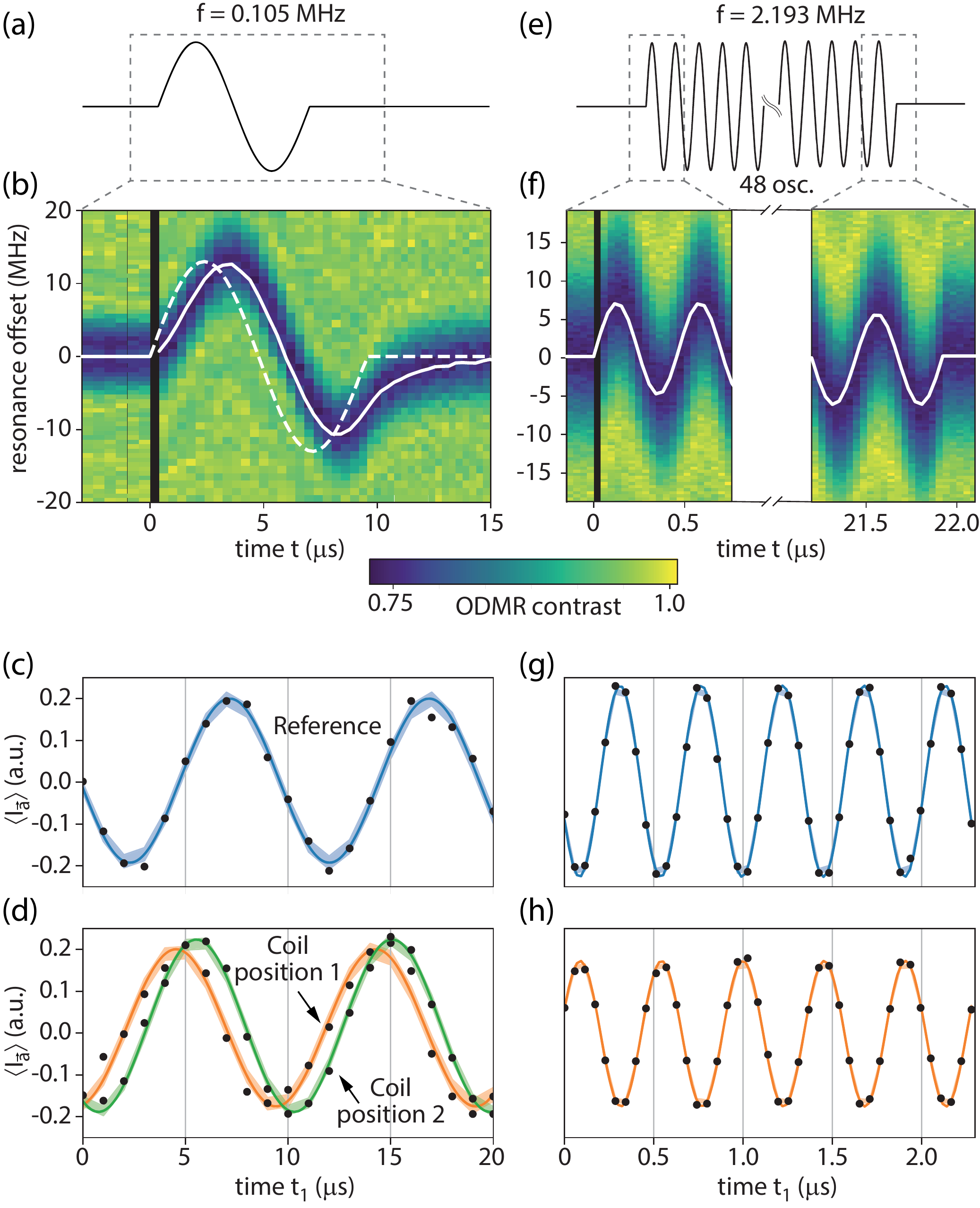}
	\caption{
(a-d) Precision measurement of the azimuth angle of \Cn{1} at low magnetic field, $\Bo=9.600(8)\unit{mT}$.
(a) Waveform of the pulse sent to the rf coil.
(b) ODMR spectra (vertical axis) of the rf coil magnetic field recorded in time steps of $400\unit{ns}$ (horizontal axis).
The black vertical line marks the start time $t=0$ of the rf pulse.
The white solid line connects the resonance positions determined by Lorentzian fits.
For comparison, we also plot the input waveform from (a) (white dashed line).
(c,d) Nuclear precession signal measured as a function of $t_1$.
Dots show the experimental data.  Colored lines represent density matrix simulations (best fit) discussed in the text.  Shaded areas specify $2\sigma$ confidence intervals of the fits.
Panel (c) shows the reference measurement (sequence of Fig. \ref{fig:fig1}(c)) and panel (d) measurements for two different coil positions (sequence of Fig. \ref{fig:fig1}(d)).
(e-h) Same experiment performed on \Cn{3} at high magnetic field, $\Bo=204.902(9)\unit{mT}$.
}
\label{fig:fig2}
\end{figure}

The key component of our experiment is the external rf coil, whose field orientation serves as the spatial reference for the $\phi$ angle measurement.  Two generations of micro-coils are used: the first coil has a 3-dB-bandwidth of $77\unit{kHz}$ (deduced from the step response recorded with the NV center) and is used for low field experiments.  The second coil reaches a bandwidth of $1.72\unit{MHz}$.  Both rf coils produce fields of $\sim 5\unit{mT/A}$ and are operated with currents of up to $1\unit{A}$. 
Crucial for our experiments is a precise knowledge of the direction and temporal shape of the coil magnetic field.  We determine the three-dimensional vector of the coil magnetic field $\vecBcoil$ using two other nearby NV centers with different crystallographic orientations with an uncertainty of less than $15\unit{\uT}$ in all three spatial components \cite{steinert10,zopes18}.  We align our ($x$,$y$,$z$) laboratory reference frame to the ([$1\bar{1}2$],[$\bar{1}10$],[111]) crystallographic axes of the single crystal diamond chips (up to an inversion symmetry about the origin).
To calibrate the dynamic response of the coil, we perform \textit{in situ} measurements of the rf field using time-resolved optically-detected magnetic resonance (ODMR) spectroscopy (Fig. \ref{fig:fig2}(a,e)).  We acquire ODMR spectra in snapshots of $400\unit{ns}$ (a) or $100\unit{ns}$ (e) over the duration of the rf pulse, and determine the pulse profile by fitting the peak positions of the resonance curves.

%


In Fig. \ref{fig:fig2}(c,d), we show a first set of measurements for nuclear spin \Cn{1} carried out at low magnetic field, $\Bo\sim 10\unit{mT}$.  The hyperfine coupling parameters of this nuclear spin are $(\apar,\aperp)=2\pi\times (18.5(1)\unit{kHz},41.4(2)\unit{kHz})$, calibrated by a separate correlation spectroscopy measurement \cite{boss16}.  Fig. \ref{fig:fig2}(c) shows the reference measurement of the nuclear spin precession after application of the $\pihalf$ pulse using the hyperfine field.
Fig. \ref{fig:fig2}(d) plots the corresponding precession signal after applying the $\pihalf$ rotation with the rf coil.
We observe a clear phase shift $\Dphi$ between the two signals, indicating that the hyperfine field $\vecAz/\yn$ and the coil field $\vecBcoil$ point in different spatial directions.  We verify that the phase shift changes if we vary the direction of $\vecBcoil$ by moving the rf coil to a different position (green data in Fig. \ref{fig:fig2}(d)).

For ideal rf pulses and exact timings, the observed phase shift $\Dphi$ corresponds to the difference $\phi-\phicoil$ between the azimuth angles of the hyperfine and coil magnetic fields, allowing us to directly deduce $\phi$.  However, due to the limited bandwidth of the rf circuit and the finite length of feed lines, the actual rf pulses tend to be delayed and distorted, leading to a phase offset.  In addition, the AC magnetometry measurement is very sensitive to timing errors and resonance offsets in the microwave modulation, causing additional uncertainty in the phase measurement.
To compensate for these issues, we determine $\phi$ by fitting the experimental data with a Levenberg-Marquardt algorithm using a density matrix simulation \cite{johansson13} as fit function and $\phi$ as fit parameter.  We propagate the two-spin density matrix through the full sequence shown in Fig. \ref{fig:fig1}(b) using piece-wise constant Hamiltonians for the nuclear spin propagation, taking the calibrated vector field and temporal shape of Fig. \ref{fig:fig2}(a) as well as the hyperfine parameters ($\apar$,$\aperp$) as inputs.  By calculating the nuclear spin evolution in the laboratory frame of reference, the simulation captures the Bloch-Siegert shift \cite{abragam61} and the $z$-component of the rf field.  In addition, we directly retrieve the absolute laboratory frame azimuth $\phi$ rather than the relative $\Dphi$ between $\vecAz$ and $\vecBcoil$.

\begin{figure}[t!]
	\includegraphics[width=0.43\textwidth]{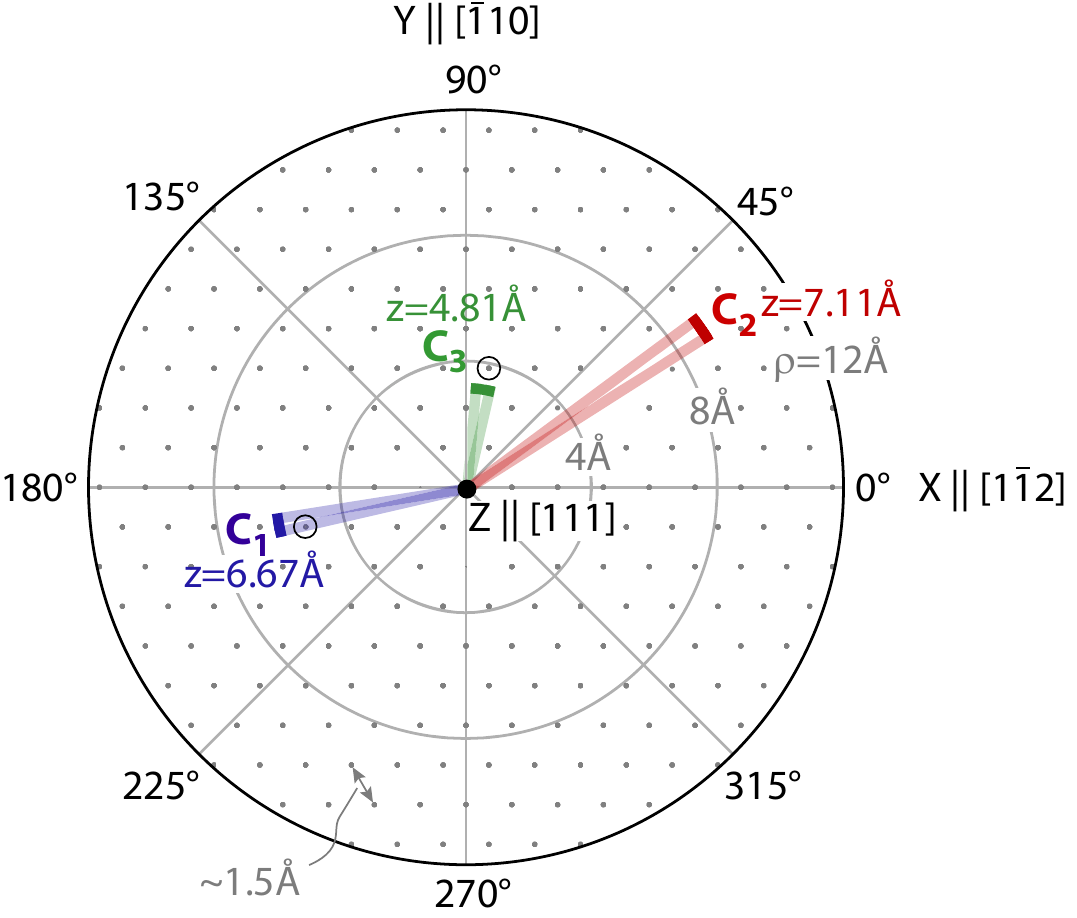}
	\caption{Polar plot of the reconstructed nuclear spin positions in the $xy$-plane of the laboratory frame.  Shaded regions mark the uncertainty in $\phi$ of the respective nuclear spin.  Radial distances $\rho=r\sin\theta$ and vertical heights $z=r|\cos\theta|$ of the nuclear sites are determined from the parallel and perpendicular hyperfine parameters by inverting the point-dipole formula \cite{zopes18}.
The measurement uncertainties in $z$ and $\rho$, neglecting deviations from the point-dipole model \cite{gali08,nizovtsev18,zopes18}, are less than $0.02\unit{\angstrom}$ for all nuclei. Grey points represent the lattice sites of diamond.  \Cn{1} and \Cn{3} are in good agreement with sites C47 and C390 (black circles) of a recent density functional theory (DFT) simulation \cite{nizovtsev18} (\Cn{2} is not part of the simulation).  The offset between experimental and best-fitting DFT locations is due to the extended NV wave function that limits the point-dipole approximation \cite{zopes18}.  The through-space distance of \Cn{2} is $r=11.5\unit{\angstrom}$.
	}
		\label{fig:fig3}
\end{figure}

We start the analysis by fitting the simulation to the reference measurement (Fig. \ref{fig:fig2}(c)), which allows us to determine $B_0$ with an uncertainty smaller than $10\unit{\uT}$. As $B_0$ defines the nuclear precession frequency, this calibration is of paramount importance for a precise estimate of $\phi$. Afterwards we determine $\phi$ with a second fit to the measurements with the rf pulse (Fig. \ref{fig:fig2}(d)) while keeping $B_0$ fixed. All fit results are shown by solid lines in Fig. \ref{fig:fig2}(c,d).  We find an azimuth location of $\phi = 191 \pm 2 \,^{\circ}$.
We have previously determined the three-dimensional coordinates of the same nuclear spin using a different positioning method \cite{zopes18}, where $\phi = 197 \pm 4 \,^{\circ}$, in good agreement with the present result.
The accuracy of our experiment is presently limited by the calibration uncertainty of the coil field angle ($\sim 1\,^{\circ}$) and by the statistical fit error of the precession phase ($\sim 1\,^{\circ}$). Additional sources of uncertainty, \eg, a misalignment of $\Bo$ or the influence of the local chemical environment are not included in the analysis, but are expected to be insignificant for our study.  The estimated three-dimensional location for this (\Cn{1}) and another nuclear spin (\Cn{2}; $(\apar,\aperp)=2\pi\times (1.9(1)\unit{kHz},19.2(1)\unit{kHz})$) are shown in Fig. \ref{fig:fig3}.
%

Next, we demonstrate that our azimuth positioning technique can be readily extended to high magnetic fields.  High bias fields are desirable in NMR because of a better peak separation and a simplified interpretation of spectra.  In addition, in NV-NMR, more efficient dynamical decoupling control and repetitive readout schemes become possible at higher fields \cite{neumann10science}.  In Fig. \ref{fig:fig2}(e-h) we show measurements carried out at $\sim 200\unit{mT}$ on a third nuclear spin (\Cn{3}) with hyperfine coupling parameters $(\apar,\aperp)=2\pi\times (98.4\unit{kHz},138.4\unit{kHz})$.  Here, we find $\phi = 81 \pm 4^{\circ}$. The three-dimensional location of \Cn{3} is also indicated in Fig. \ref{fig:fig3}.

The $\phi$ uncertainty at high magnetic field is larger than at low field because of timing errors.  At $200\,\unit{mT}$, the nuclear Larmor period is only $\sim 460\unit{ns}$, such that $1\unit{ns}$ of timing uncertainty causes a phase uncertainty of about $0.8^\circ$.  For the rf pulse in Fig. \ref{fig:fig2}(e), we find a phase delay of $12\pm 3\,^{\circ}$, corresponding to an overall timing uncertainty of the ODMR calibration of $\sim 4\unit{ns}$.  Although the measured phase delay is in good agreement with the value predicted from the electrical characteristics of the rf circuit ($\sim 11^\circ$), it already introduces the largest error to the $\phi$ measurement.
For future experiments carried out in the high bias fields of superconducting magnets \cite{aslam17} a precise calibration of control fields will therefore become even more critical.

\begin{figure}[]
\includegraphics[width=0.48\textwidth]{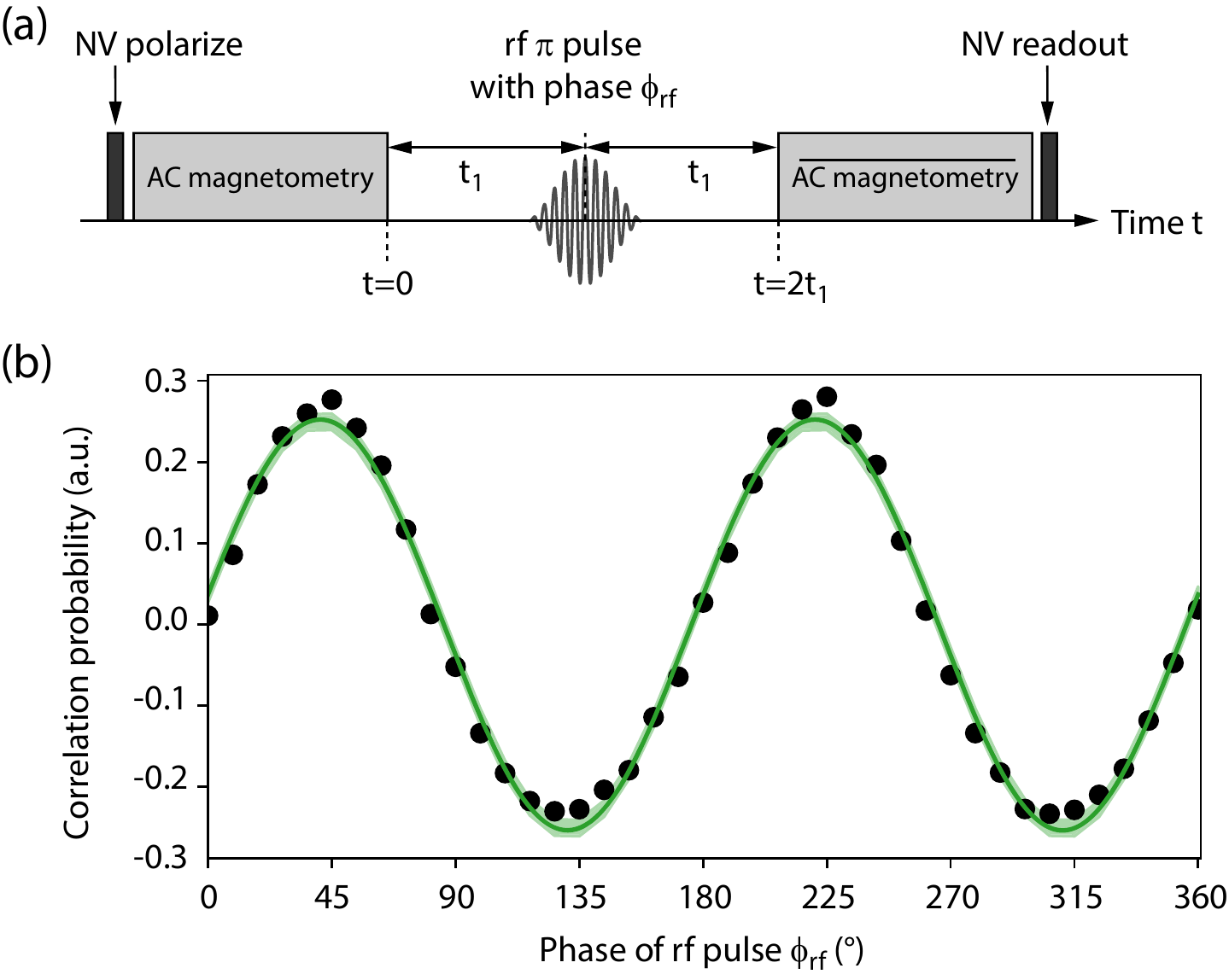}
\caption{Measurement of the hyperfine $\phi$ angle by a nuclear spin echo.
(a)	Pulse sequence of the experiment: The free evolution time of a correlation spectroscopy sequence is interspersed with a $\pi$ pulse generated by the rf coil.  A cosine-square envelope \cite{zopes17} is used to suppress pulse transients, and the pulse is selective to the nuclear spin transition associated with the electronic $\mS=0$ state.  The correlation spectroscopy sequence is implemented by two AC magnetometry blocks as in Fig. \ref{fig:fig1}(e); the bar on the second block indicates that the sequence is reversed.  
(b) Spin echo modulation detected on \Cn{3}.  Black dots show the data and the green line shows the density matrix simulation (best fit with $\phi$ as free parameter).  The $2\sigma$ confidence intervals of the fit are indicated by shaded areas.  The evolution time is $2t_1 = 31.36\unit{\us}$.
}
\label{fig:fig4}
\end{figure}

Finally, we discuss a complementary scheme for reconstructing the azimuth angle that does not require pre-polarization of nuclear spins.  Instead of recording the nuclear precession signal as a function of $t_1$, we intersperse a correlation spectroscopy sequence \cite{laraoui13,boss16} with a central rf $\pi$ pulse to generate a nuclear spin echo at a fixed time $t=2t_1$ (Fig. \ref{fig:fig4}(a)).  By varying the pulse phase $\phirf$ from $0-360^\circ$, we modulate the amplitude of the spin echo, leading to an oscillatory signal $\propto \cos(2\phirf-2\phi)$.  We then determine $\phi$ from the phase offset of the oscillation.
Fig. \ref{fig:fig4}(b) shows a spin echo oscillation for \Cn{3} measured at a bias field of $204.9(1)\unit{mT}$.  The compatible angles are $\{88\pm4^\circ,268\pm4^\circ\}$, in good agreement with the result from Fig. \ref{fig:fig2}(h).  Note that the echo method is afflicted by a $180^\circ$ ambiguity in the angle measurement, because the echo oscillation repeats with $\phirf$ modulo $\pi$.  Although the ambiguity could possibly be resolved by applying concomitant rf and microwave rotations or by introducing dc field pulses \cite{zopes18}, it is unlikely to restrict future experiments on single molecules where relative, rather than absolute, positions are important.  In addition, single-molecule NMR experiments can exploit internuclear interactions to further constrain the nuclear positions.
%
%


In conclusion, we have introduced a simple method for measuring the inter-spin azimuth $\phi$, enabling us to perform three-dimensional distance measurements on single nuclear spins.  We demonstrate the potential of our technique by mapping the 3D location of individual \C nuclei in diamond with a precision sufficient for assigning discrete lattice sites.
Future experiments will apply 3D distance measurements to molecules deposited on the surface of dedicated diamond NMR sensor chips \cite{mamin13,ohashi13,loretz14apl,lovchinsky16} and provide an avenue to analyze the structure and conformation of single molecules with atomic resolution \cite{sidles09}.


This work was supported by Swiss National Science Foundation (SNFS) Project Grant No. 200020\_175600, the National Center of Competence in Research in Quantum Science and Technology (NCCR QSIT), and the DIAmond Devices Enabled Metrology and Sensing (DIADEMS) program, Grant No. 611143, of the European Commission.
We thank A. Nizovtsev and F. Jelezko for sharing details about the DFT simulation.  While finishing this manuscript, we learned about a similar idea put forward by Sasaki and coworkers \cite{sasaki18}.\\

\vspace{0.0cm}\noindent
{\small $^\dagger$These authors contributed equally to this work.}\\
{\small $^\ast$Email: \href{mailto:degenc@ethz.ch}{degenc@ethz.ch}}

\input{"references.bbl"}

\end{document}

%% file: references.bbl
%